\documentclass[12pt]{iopart}
\usepackage{comment}
\usepackage[utf8]{inputenc}
\usepackage{graphicx}
\usepackage{tikz}
\usetikzlibrary{decorations.pathmorphing}
\usepackage{subcaption}
\usepackage{caption}
\expandafter\let\csname equation*\endcsname\relax
\expandafter\let\csname endequation*\endcsname\relax
\usepackage{amsmath}

\begin{document}

\title{An improved car-oriented mean-field theory for stochastic traffic flow models}

\author{Yasar Efe Dai$^{1,a}$, Andreas Schadschneider$^{1,3,b}$, Michael Schreckenberg$^{2,c}$}

\address{$^1$Institut f\"ur Theoretische Physik, Universit\"at zu K\"oln,
50937 K\"oln, Germany}
\address{$^2$Physik von Transport und Verkehr, Universit\"at Duisburg-Essen, 47048 Duisburg, Germany}
\address{$^3$Institut f\"ur Physikdidaktik, Universit\"at zu K\"oln,
50931 K\"oln, Germany}
\ead{$^a$ydai2@smail.uni-koeln.de, $^b$as@thp.uni-koeln.de,\\$^c$michael.schreckenberg@uni-due.de}
\vspace{10pt}
\begin{indented}
\item[]\today
\end{indented}

\begin{abstract}
We propose an improved mean-field analysis of cellular automata models of single-lane vehicular traffic. 
By combining aspects of the Car-Oriented-Mean-Field (COMF) theory and the 2-site cluster method, which have been previously successfully applied to similar models, we aim to capture both short- and long-range correlations more accurately. 
In contrast to classical mean-field theories, 
the improved method is well suited for models with inhomogeneous stationary states and able to capture the essential properties of phase separation, e.g. in models with slow-to-start rules. The improved accuracy and new physical insights are illustrated through an application to the VDR model with  $v_{\text{max}}=1$.
\end{abstract}

\section{Introduction}

Cellular automata (CA) models are a very useful tool for the simulation of highway and urban traffic  \cite{Chowdhury2000,SCN-Book}. Due to their discrete space-time nature and rule-based dynamics, they are ideally suited for large-scale computer simulations. This makes faster-than-real-time simulations of large highway networks possible and allows for traffic flow predictions (see e.g.\ \cite{EsserS97,NagelER00,WahleNES01} for early works). Nevertheless, exact analytical results can provide deeper insight into the basic mechanisms of such models, for example, the occurrence of phase transitions \cite{ChowdhuryKNSS00,RotersLU00} or the existence of metastable states.

The \emph{Nagel-Schreckenberg (NaSch)} model introduced in~\cite{NaSch} is a seminal model for traffic flow. It is closely related to the \emph{Asymmetric Simple Exclusion Process (ASEP)}. 
The ASEP is the paradigmatic model for driven diffusion in one dimension and has been studied intensively so that several exact results are known \cite{DERRIDA199865,Schuetz-review,EvansB02}. The special case $v_\text{max}=1$ of the NaSch model is equivalent to the \emph{Totally Asymmetric Simple Exclusion Process (TASEP)} with parallel dynamics where all lattice sites are updated simultaneously and particles are allowed to diffuse only in one direction.  

In contrast to the continuous time TASEP, where site occupations are uncorrelated and a simple mean-field approach is exact for the stationary state in the case of periodic boundary conditions, the TASEP with parallel dynamics exhibits strong short-range correlations. These can be attributed to the existence of so-called \emph{Garden of Eden states (GOE)} \cite{SchadschneiderS98}, i.e.\ states that cannot be reached by the dynamics. These states cannot be captured by mean-field theory, but by the \emph{two-site cluster} approach \cite{SchreckenbergSNI95} and the \emph{Car-Oriented-Mean-Field (COMF)} theory \cite{SchadschneiderS97} which yield the exact stationary state.

In this work, we investigate a novel analytical approach that builds on the COMF theory and incorporates elements of the two-cluster approach. The new description aims at improving the accuracy of the well-known COMF theory to better capture the underlying physics of CA models for traffic flow. We present it in the context of an extended NaSch model, the \emph{Velocity-Dependent-Randomization (VDR)} model \cite{Barlovic_1998}. The VDR model is a natural generalization of the NaSch model where the slowing down probability depends on the momentary velocity of the car.

In COMF theory, the state of the system is described by the velocities of the vehicles and the numbers of empty sites in front of them, usually referred to as the headway. It is a genuine mean-field theory in which all correlations between the dynamical variables of different vehicles are neglected. Within this framework, the distributions $P_v(n)$ of vehicles with velocity $v$ and headway $n$ are treated exactly\footnote{COMF can be interpreted by mapping of the TASEP to a zero-range process (ZRP) \cite{Evans_2005}, which admits a factorized steady state \cite{Spitzer}.}.
In the two-cluster approach, the state of the system is described by the occupation numbers of pairs of sites.
The two-site probabilities $P_{\sigma,\sigma'}$ are treated exactly, where $\sigma$ and $\sigma'$ denote the occupation numbers of some neighbouring sites.

Although exact for the NaSch model with $v_{\text{max}}=1$, COMF fails to accurately describe inhomogeneous stationary states such as those observed in models with slow-to-start (STS) rules \cite{Schadschneider_1997} like the Takayasu–Takayasu model \cite{TakayasuT93a}, Benjamin–Johnson–Hui model \cite{BenjaminJH96} and VDR model \cite{Barlovic_1998}.  
Here, for intermediate and high densities, the system divides into two distinct phases: a large compact jam with dominantly short headways, and a free-flow region with large headways. Within the jammed state, one observes strong short-range correlations due to the higher likelihood that the site adjacent to a standing car is also occupied by a stationary vehicle.  In these cases, COMF becomes only an approximation \cite{Schadschneider_1997} since it would incorrectly assume that the headways are randomly distributed. 

The simplest approach to capture correlations between neighboring cars, relevant in this case, is arguably incorporating the velocity of the vehicle immediately ahead. This allows for accounting for some short-range correlations. In the following we develop such an approach which will then be applied to the VDR model with $v_{\text{max}}=1$ and parallel dynamics.


\section{The Velocity-Dependent-Randomization Model} 

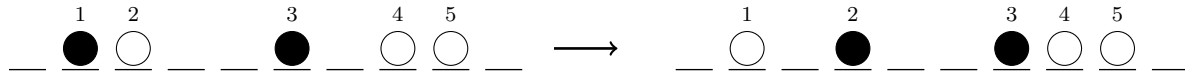
\begin{figure}[h!]
\label{fig: Model Illustration}
\centering
\begin{tikzpicture}[scale=0.7]
  \begin{scope}[xshift=0cm]
    \foreach \i in {1,...,10} {
      \draw[line width=0.5pt] (\i,0) -- (\i+0.7,0);
    }
    \node at (2.35, 1.05) {\scriptsize 1};
    \node at (3.35, 1.05) {\scriptsize 2};
    \node at (6.35, 1.05) {\scriptsize 3};
    \node at (8.35, 1.05) {\scriptsize 4};
    \node at (9.35, 1.05) {\scriptsize 5};
    \filldraw[black] (2.35,0.4) circle (0.32);          
    \draw[black, fill=white] (3.35,0.4) circle (0.32);  
    \filldraw[black] (6.35,0.4) circle (0.32);          
    \draw[black, fill=white] (8.35,0.4) circle (0.32);  
    \draw[black, fill=white] (9.35,0.4) circle (0.32);  
  \end{scope}
  \draw[->, line width=1pt] (11.3, 0.4) -- (12.5, 0.4);
  \begin{scope}[xshift=12.6cm]
    \foreach \i in {1,...,10} {
      \draw[line width=0.5pt] (\i,0) -- (\i+0.7,0);
    }
    \node at (2.35, 1.05) {\scriptsize 1};
    \node at (4.35, 1.05) {\scriptsize 2};
    \node at (7.35, 1.05) {\scriptsize 3};
    \node at (8.35, 1.05) {\scriptsize 4};
    \node at (9.35, 1.05) {\scriptsize 5};
    \draw[black, fill=white] (2.35,0.4) circle (0.32);  
    \filldraw[black] (4.35,0.4) circle (0.32);          
    \filldraw[black] (7.35,0.4) circle (0.32);          
    \draw[black, fill=white] (8.35,0.4) circle (0.32);  
    \draw[black, fill=white] (9.35,0.4) circle (0.32);  
  \end{scope}
\end{tikzpicture}
\caption{A parallel update of the VDR model on a lattice of length $L=10$ with $N=5$ cars in the limit $p=0$, $p_0>0$. Black (white) circles denote cars with velocity $1$ ($0$). The configuration at time $t$ (left) leads to the configuration at time $t+1$ (right) if car 2 stochastically restarts and car 5 stochastically remains stationary.}
\label{fig:vdr-update}
\end{figure}

We consider a one-dimensional lattice with $L$ sites and $N$ vehicles. The density $\rho = N/L$ of vehicles is conserved for the case of periodic boundary conditions. The original NaSch model is defined by the set of four rules, (NaSch~1)--(NaSch~4), which are applied in order and synchronously to all vehicles in a given configuration at time $t$ to obtain the next configuration at time $t+1$. The VDR model extends this framework by introducing an additional rule, (VDR~0), in which the braking probability assigned to a vehicle depends on its velocity at the end of the previous update \cite{Barlovic_1998}. This assigned probability is then used in the subsequent random-braking step, (NaSch~3). For completeness, we list the five update rules of the VDR model with $v_{\text{max}}=1$.
\begin{itemize}
    \item[] (VDR~0) \textbf{Determination of the randomization parameter}. The random braking parameter of the car with velocity $v$ is given by 
    \begin{equation}
    p(v) = 
    \begin{cases}
        p_0 & \text{if } v = 0, \\
        p & \text{if } v = 1.
    \end{cases}
\end{equation}
    
    \item[] (NaSch\,1) \textbf{Acceleration}. The car is accelerated to velocity $1$.
    
    \item[] (NaSch\,2) \textbf{Deceleration}. If the car's headway is zero, then the car becomes stationary, i.e.\ $v=0$.
    
    \item[] (NaSch\,3) \textbf{Random-braking (randomization)}. The car is subject to stochastic braking with probability $p(v)$ determined in step (VDR~0), i.e.\ if the car has velocity $1$ after step (NaSch~2) then its velocity is reduced to $v=0$ with probability $p(v)$.
    
    \item[] (NaSch\,4) \textbf{Movement}. If the car has velocity $v=1$ after step (NaSch~3), then it moves forward by one site.
\end{itemize}
We note that for $p_0=p$, these rules are identical to those of the NaSch model. 
Fig.~\ref{fig:vdr-update} illustrates a parallel update of the VDR model with $v_{\text{max}}=1$ and with $p=0$, $p_0>0$.


\section{The improved Car-Oriented-Mean-Field (iCOMF) Theory}
\label{Sec: iComf Theory}

Instead of a conventional site-oriented description of the system configurations, based on occupation numbers as the dynamical variables, we adopt here a car-oriented description based on the headway $n_i(t)$, i.e.\ the number of empty cells in front of car $i$, and its velocity $v_i(t)$  which fully describes the configuration of the system\footnote{More precisely, the position for one car has to be tracked for a full mapping. However, the stationary state for periodic boundary conditions is translational invariant so that this information is not essential.}.
We recall that the COMF theory ignores correlations between the velocities and headways of cars in front, assuming a factorized steady state of the form 
\begin{equation}    
P(\{n_i; v_i\}) \sim \prod_{i=1}^{N} P(n_i; v_i).
\end{equation}
For the VDR model with $v_{\max}=1$ and general braking probabilities $p_0$ and $p$, COMF theory is not exact \cite{Barlovic-Diploma}.
Therefore we extend it by additionally incorporating the velocity of the vehicle immediately ahead. This allows the theory to capture short-range correlations between neighbouring vehicles while retaining the analytical structure of the COMF theory. The improved COMF theory (iCOMF) can be represented as a product state over nearest-neighbor pairs of cars
\begin{equation}
P(\{n_i; v_i\}) \sim \prod_{i=1}^{N} P(n_i; v_i,v_{i+1}),
\end{equation}
and treats consecutive pairs of cars $i$ and $i+1$, with velocities $v_i$ and $v_{i+1}$, separated by a distance $n_i$, in an exact manner. These pairs are then coupled to the rest of the system in a self-consistent fashion. To achieve this, one needs to formulate the master equation for the probability distribution of finding exactly $n$ empty sites between a car with velocity $u$ and the car immediately ahead with velocity $v$. This probability is denoted by $P(n; u,v,t) = P_{uv}(n,t)$. We introduce the conditional probabilities $g_v$ $(\bar{g}_v)$ that a car moves (does not move) at the end of the update started at time t, given that its velocity at time $t$ is $v=0,1$:
\begin{align}
\label{g_v}
    g_v(t) 
    &= \frac{\big(1-p(v)\big)}{P_v(t)} \sum_{n=1}^{\infty} \sum_{v'=0,1} P_{v v'}(n, t) 
    	= \Big(1-p(v)\Big) \left( 1 - \frac{P_{v0}(0,t)}{P_v(t)} \right), \\
    \bar{g}_v(t) &= 1 - g_v(t),
\end{align}
where $P_v(t)$ is the probability of finding a car with velocity $v$ at time $t$ 
\begin{equation}
    P_v(t) = \sum_{n=0}^{\infty} \sum_{v'=0,1} P_{vv'}(n, t). 
    \label{eq:pv}
\end{equation}
\begin{figure}[h!]
\centering
\begin{tikzpicture}[scale=0.7]
  \begin{scope}[xshift=0cm]
    \draw[line width=0.5pt] (2,0) -- (2.7,0);
    \draw[line width=0.5pt] (3,0) -- (3.7,0);
    \node at (4.35, 0) {$\cdots$};
    \draw[line width=0.5pt] (5,0) -- (5.7,0);
    \draw[line width=0.5pt] (6,0) -- (6.7,0);
    \draw[black, fill=white] (2.35,0.4) circle (0.32) node {\scriptsize $u$};
    \draw[black, fill=white] (6.35,0.4) circle (0.32) node {\scriptsize $v$};
    \draw[->, line width=0.5pt] (2.35,0.75) to[bend left=70] (3.35,0.75);
    \draw[->, line width=0.5pt] (6.35,0.75) to[bend left=70] (7.35,0.75);
    \draw[red, line width=0.7pt] (6.65,0.85) -- (7.05,1.20);
    \draw[red, line width=0.7pt] (7.05,0.85) -- (6.65,1.20);
    \node at (2.85, 1.6) {\scriptsize $1-p(u)$};
    \node at (6.85, 1.6) {\scriptsize $\bar{g}_v$};
    \node at (4.35, -0.55) {\scriptsize $n+1$};
  \end{scope}
  \draw[->, line width=1pt] (8.4, 0.4) -- (9.6, 0.4);

  \begin{scope}[xshift=9.3cm]
    \draw[line width=0.5pt] (1,0) -- (1.7,0);
    \draw[line width=0.5pt] (2,0) -- (2.7,0);
    \draw[line width=0.5pt] (3,0) -- (3.7,0);
    \node at (4.35, 0) {$\cdots$};
    \draw[line width=0.5pt] (5,0) -- (5.7,0);
    \draw[line width=0.5pt] (6,0) -- (6.7,0);
    \filldraw[black] (2.35,0.4) circle (0.32);
    \draw[black, fill=white] (6.35,0.4) circle (0.32);
    \node at (4.35, -0.55) {\scriptsize $n$};
  \end{scope}
\end{tikzpicture}
\caption{Schematic of the master equation for $P_{10}(n,t+1)$. It evolves from $P_{uv}(n+1,t)$ if the leading car does not move (with probability  $\bar{g}_v$) and the following car moves one step (with probability $1-p(u)$). This is the only transition leading to $P_{10}(n,t+1)$.}
\label{fig:master-equation-representation}
\end{figure}
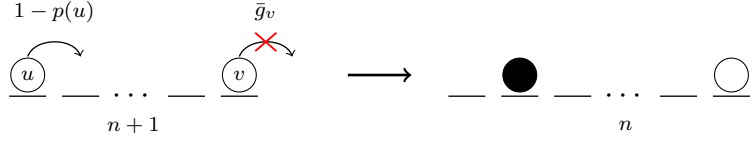
A schematic representation of the master equation governing the time evolution of $P_{10}(n,t)$ is shown in Fig.~\ref{fig:master-equation-representation}. 
A configuration at time $t+1$ (right) in which the following car has velocity $1$ and headway $n$, while the leading car has velocity $0$, can only arise from configurations at time $t$ (left) where the following car has velocity $u=0,1$ and headway $n+1$, and the leading car has velocity $v=0,1$. The transition occurs when the following car moves with probability $1-p(u)$ and the leading car remains stationary with probability $\bar g_v$, which leads to
\begin{equation}
        P_{10}(n,t+1) = \sum_{u=0,1}\sum_{v=0,1} \big(1-p(u)\big) \bar{g}_v(t)P_{uv}(n+1,t) ,\quad n \geq 0.
\end{equation}  
In the stationary state, the probabilities $P_{u,v}(n,t)$ become time-independent and are obtained by solving an infinite set of nonlinear master equations. The details of the calculation are provided in~\ref{Appendix-iCOMFdetails}.
Throughout this work we compare our results with those obtained using the original COMF theory. A summary of this calculation can be found in \ref{sec: Original COMF}.


\section{The Cruise-Control Limit and Phase-Separated States}

\begin{figure}[h]
    \centering
    \includegraphics[width=0.9\linewidth]{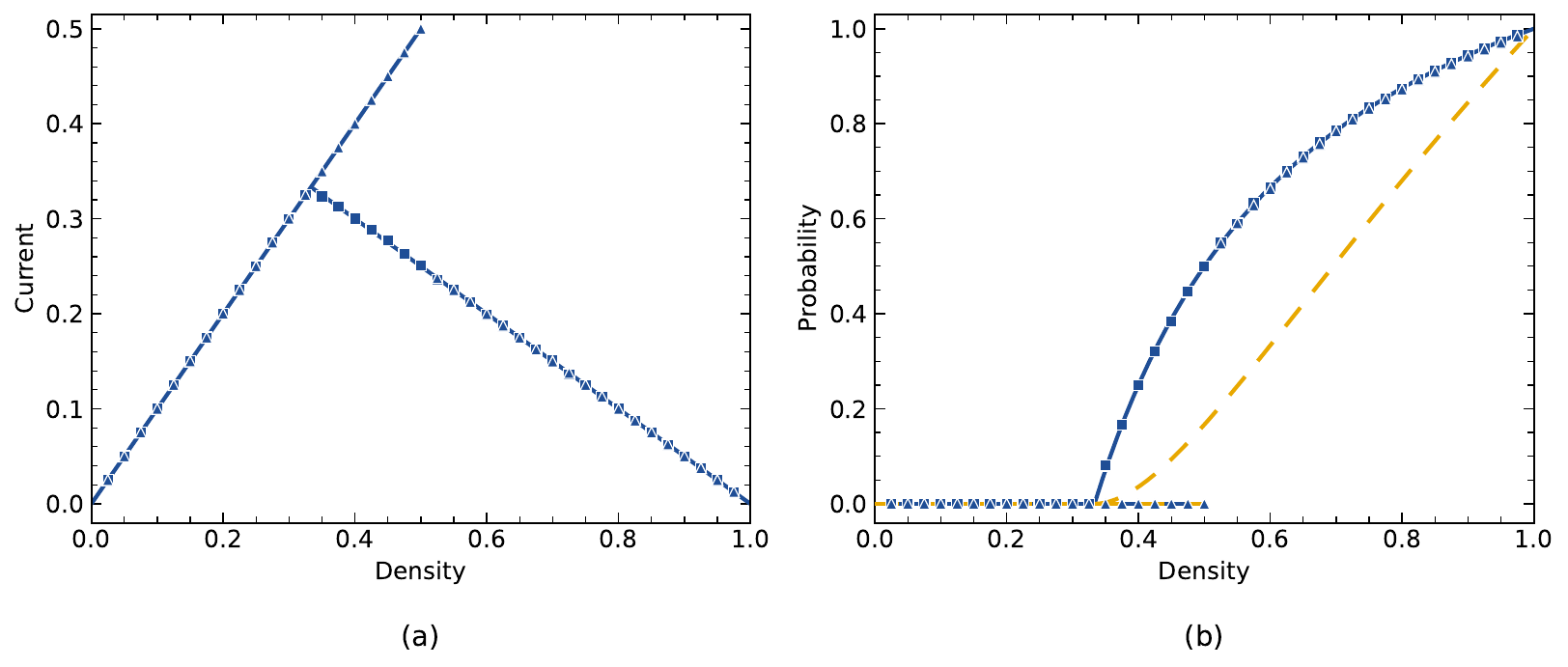}
    \caption{(a) Fundamental diagram and (b) zero-headway distribution of a stationary car with $p_0=0.5$ and $p=0$. The triangular (rectangular) markers denote simulations data with homogeneous (jammed) initial conditions. They are compared with stable and metastable theoretical results. In (a) COMF and iCOMF results coincide and are represented as continuous line.  In (b) COMF and iCOMF results are represented by a yellow dashed line and blue continuous line, respectively. They only coincide in the free-flow regime.}
    \label{ZeroHeadway-Model0}
\end{figure}

\begin{figure}[h]
    \centering
    \includegraphics[width=0.9\linewidth]{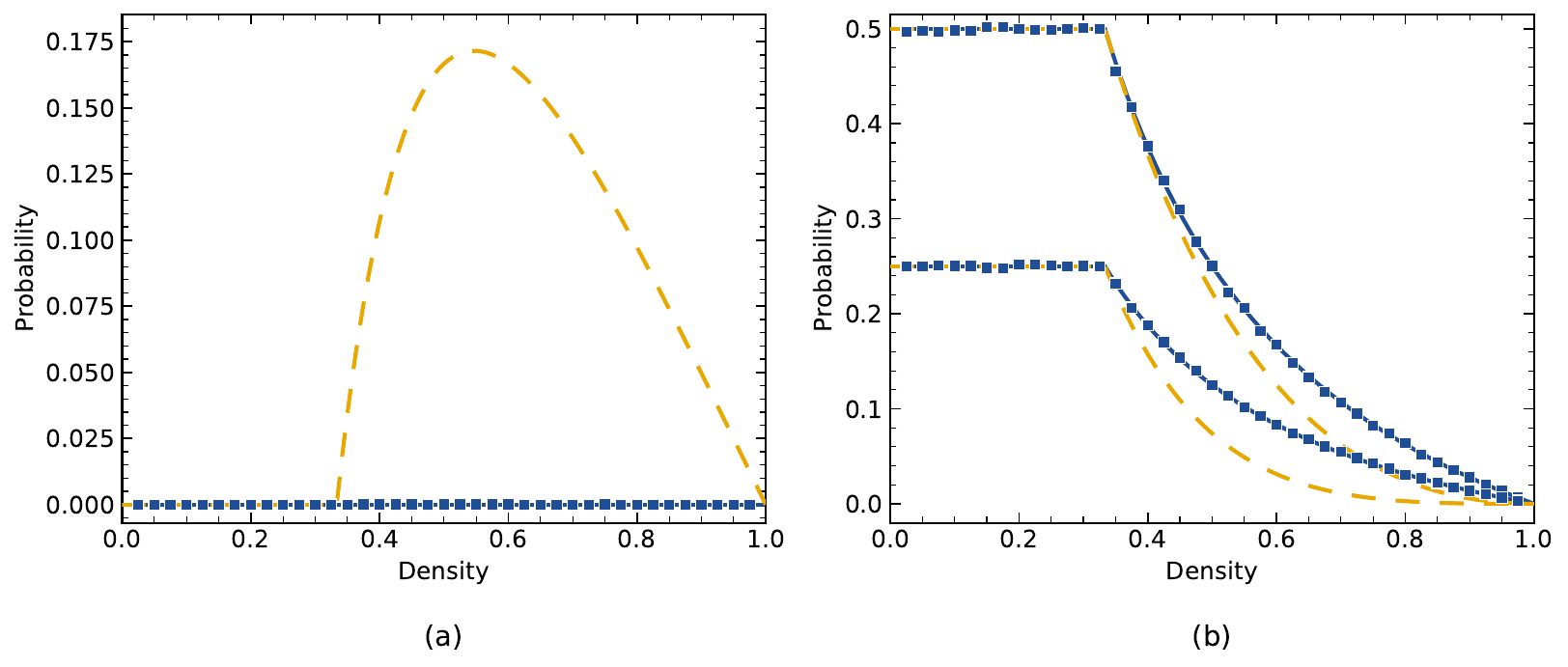}
    \caption{Headway distribution of a moving car in the cruise-control limit with
    $p_0=0.5$, $p_1=0$: (a) 0-headway distribution and (b) non-zero headway distributions (1-headway and 2-headway from top to bottom) are shown. Computer simulations with a jammed initial configuration (markers) are compared with COMF (yellow dashed lines) and iCOMF results (continuous lines).}
    \label{fig:Headway11_CC}
\end{figure}

For traffic flow applications, one is mainly interested in the slow-to-start (STS) regime where $p_0>p$. A particularly interesting case is when  the a moving car can only slow down in the deceleration step (NaSch~2), i.e.\ due to insufficient headway, and not by randomization i.e. $p=0$ and $0 < p_0 < 1$. We note that for $v_{max}=1$, this limit of the VDR model corresponds to the \emph{cruise-control limit} of the NaSch model \cite{CruiseControl}. In this regime, the fundamental diagram has the so-called \emph{inverse lambda shape} and metastable states are observed in computer simulations within a well defined density interval $\rho_1 < \rho \leq \rho_2$, where the steady state current is not uniquely determined and depends on the initial configuration of the system. The flow-density relationship determined with COMF theory \eqref{comf current}  and with iCOMF theory \eqref{eq:iCOMF_Current} coincide and is given explicitly by
\begin{equation}
\label{eq: FD_Model0}
   J(\rho) = \begin{cases}
       \rho & \text{if } 0\leq \rho \leq 1/2 \\
       (1-p_0)(1-\rho) & \text{if } \frac{1-p_0}{2-p_0}\leq \rho \leq 1
   \end{cases}.
\end{equation}
This expression includes the metastable free-flow branch and allows us to identify $\rho_1=\frac{1-p_0}{2-p_0}$ and $\rho_2=\frac{1}{2}$. The result \eqref{eq: FD_Model0} is in very good agreement with simulation data as shown in Fig.~\ref{ZeroHeadway-Model0}a. In this limit, the COMF result for the zero-headway distributions $P_v(0)^{\text{(COMF)}}$, i.e.\ the probability of finding exactly zero empty sites in front of car with velocity $v=0,1$ are given by
\begin{align}
\label{eq:COMF_Headway0_v_0}
P_{0}(0)^{\text{(COMF)}}
&= \frac{\big(2\rho-1+p_0(1-\rho)\big)^2}{\rho(\rho+p_0(1-\rho))} \,
\theta\!\left(
\rho-\frac{1-p_0}{2-p_0}
\right), \\
P_{1}(0)^{\text{(COMF)}} \label{eq:COMF_Headway0_v_1}
&= \frac{(1-p_0)(1-\rho)}{2\rho-1+p_0(1-\rho)} P_{0}(0)^{\text{(COMF)}},
\end{align}
where the Heaviside function $\theta$ ensures that the expression vanishes below $\rho_1$. The non-zero headway (for $n\geq1$) distributions can also be written explicitly and are given by
\begin{align}
\label{eq:non_zer0_headways_COMF_0}
   P_0^{\text{(COMF)}}(n) &= \begin{cases}
       0 & \text{if }  \rho \leq \rho_1 \\
       \frac{2\rho-1+p_0(1-\rho)}{p_0^2(1-\rho)} \big( \frac{p_0(1-\rho)}{\rho+p_0(1-\rho)} \big)^{n+1} & \text{if } \rho_1< \rho 
   \end{cases}, \\
\label{eq:non_zer0_headways_COMF_1}
   P_1^{\text{(COMF)}}(n) &= \begin{cases}
       (1-p_0)p_0^{n-1} & \text{if }  \rho \leq \rho_1 \\
       \frac{1-p_0}{p_0^2} \big( \frac{p_0(1-\rho)}{\rho+p_0(1-\rho)} \big)^{n+1} & \text{if } \rho_1 < \rho 
   \end{cases}.
\end{align}
These results correspond to the stable solution of the master equation and are only in agreement with the simulations data in the free-flow phase $\rho\leq \rho_1$ (Fig.~\ref{ZeroHeadway-Model0}b and Fig.~\ref{fig:Headway11_CC}a) where each car has velocity 1, at least one empty site in front and moves deterministically.  The reason for this is that in the free-flow phase the model can be mapped to a usual ZRP which implies that the COMF results \eqref{eq:COMF_Headway0_v_0}-\eqref{eq:non_zer0_headways_COMF_1} and are exact for $\rho \leq \rho_1 $. 

In the congested phase, the steady state consists of a single mega-jam of stationary cars coexisting with a free-flow region, in which cars moves at velocity one with at least one empty site in front. As expected, COMF fails to account for correlations responsible for such inhomogeneous steady state configurations and underestimates the weight of stationary cars inside a jam (Fig.~\ref{ZeroHeadway-Model0}b). Another aspect that highlights the limitations of COMF in the congested phase can be seen in the distributions \eqref{eq:COMF_Headway0_v_1} and \eqref{eq:non_zer0_headways_COMF_1}. The mega-jam  is either a cluster of cars starting with a car with velocity 0 or 1 i.e.\ a local configuration the form "e0\dots 0e" or "e10\dots 0e" (here "e" denotes empty sites and "0" ("1") denotes sites occupied by a stationary (moving) car), while the free flow phase has local states in the form "e1e\dots e1e". Clearly the states "10" and "0e\dots e1" can only occur at most once in this phase separated configuration (with finite system size) and their weight must identically vanish in the thermodynamic limit.  The COMF distributions \eqref{eq:COMF_Headway0_v_1} and \eqref{eq:non_zer0_headways_COMF_1} fails to capture this. We note that it is rather interesting that COMF fails to describe the headway distributions but captures the correct fundamental diagram, which, owing to \eqref{eq:g explicit for COMF}, is expressed in terms of both zero-headway distributions alone. This partial agreement could be attributed to the fact that, in this cruise-control limit, many different configurations leads to the same flow. This behavior has also been observed previously in COMF for the NaSch model with $v_\text{max}=2$, in the deterministic limit \cite{Schadschneider_1999}.

On the other hand, using the iCOMF theory, the zero-headway distributions, together with the GOE distributions $P_{01}(0)=P_{11}(0) = 0$, can also be obtained explicitly and are given by
\begin{align}
\label{eq:iCOMF_Headway0_v_00}
P_{00}(0)^{\text{(iCOMF)}}
&= \frac{2\rho-1+p_0 (1-\rho)}{\rho} \,
\theta\!\left(
\rho-\frac{1-p_0}{2-p_0}
\right),\\
P_{10}(0)^{\text{(iCOMF)}} &=0. \label{eq:iCOMF_Headway0_v_10}
\end{align}
Furthermore, the non-zero headway distributions for all $n\geq 1$ are given by
\begin{align}
\label{eq:non_zero_headways_iCOMF}
P_{00}(n)^{\text{(iCOMF)}} &= 0,   \\ 
\label{eq:non_zero_headways_iCOMF_01}
P_{01}(n)^{\text{(iCOMF)}} &= P_{10}(n)^{\text{(iCOMF)}} = 0, \\
\label{eq:non_zero_headways_iCOMF_11}
   P_{11}(n)^{\text{(iCOMF)}} &= \begin{cases}
       (1-p_0)p_0^{n-1} & \text{if }  \rho \leq \rho_1 \\
       \frac{1-\rho}{\rho} (1-p_0)^2 p_0^{n-1} & \text{if } \rho_1 < \rho 
   \end{cases}.
\end{align}
These iCOMF headway distributions are in excellent agreement with computer simulations across all densities as shown in Fig.~\ref{ZeroHeadway-Model0}b and Fig.~\ref{fig:Headway11_CC}. We note that, as discussed above, the only configurations that survive in the congested regime are stationary cars immediately behind another stationary car, and moving cars separated from a moving car ahead by at least one empty site. All other configurations must vanish in the thermodynamic limit, a feature correctly captured by the iCOMF headway distributions.

 We note that for $\rho\leq \rho_1$, the iCOMF headway distributions \eqref{eq:iCOMF_Headway0_v_00}-\eqref{eq:non_zero_headways_iCOMF_11} recovers the exact results \eqref{eq:COMF_Headway0_v_0}-\eqref{eq:non_zer0_headways_COMF_1}. We also note that in the cruise-control limit the VDR model becomes fully deterministic throughout the free-flow phase, which leads to non-unique the stationary states. The solution given in \eqref{eq:non_zer0_headways_COMF_1} and \eqref{eq:non_zero_headways_iCOMF_11} for $\rho \leq \rho_1$ corresponds to the headway distributions that can be obtained from computer simulations starting with a jammed initial configuration. We can see in Fig.~\ref{fig:Headway11_CC} that the excellent agreement between the iCOMF headway distributions and simulation data extends also to the non-zero headway distributions \eqref{eq:non_zero_headways_iCOMF}-\eqref{eq:non_zero_headways_iCOMF_11}. We also note the average distance between cars in the free-flow region is given by $\Delta x = 1 + (1-p_0)^{-1}$~\cite{Barlovic_1998, Appert_2001}, i.e.\ the average headway is $\langle n \rangle = (1-p_0)^{-1}$. This result is not captured by the COMF distribution in \eqref{eq:non_zer0_headways_COMF_1}, but is recovered by the iCOMF $P_{11}(n)$ distribution in \eqref{eq:non_zero_headways_iCOMF_11}.


\section{The General Model ($p>0$)}

\begin{figure}
    \centering    \includegraphics[width=1\linewidth]{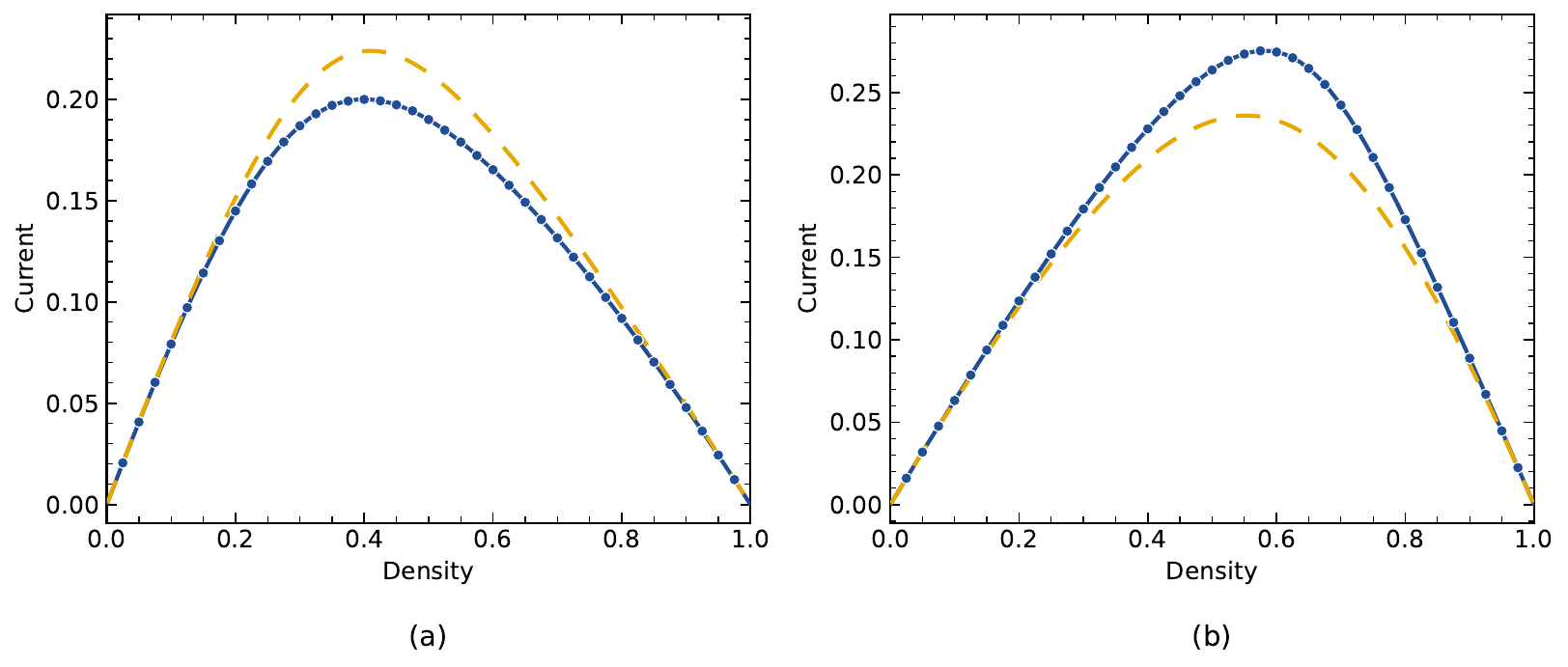}
    \caption{Generic fundamental diagrams in the STS regime with $p_0=0.5, p=0.1$ (a) and FTS regime with $p_0=0.1, p=0.5$ (b). Comparison between iCOMF (solid blue lines), COMF (yellow dotted lines) and computer simulations (markers) are shown.}
    \label{FD- General Model}
\end{figure}

\begin{figure}
    \centering
    \includegraphics[width=1\linewidth]{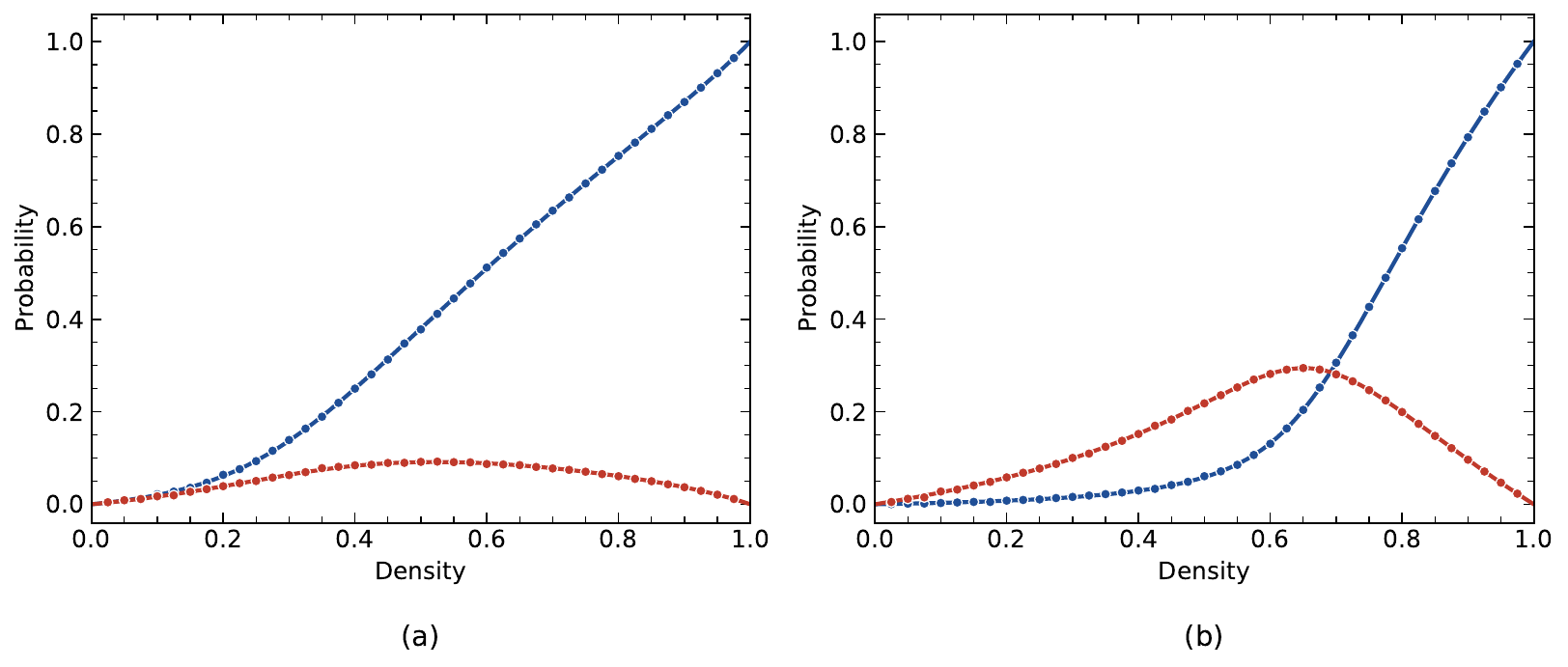}
    \caption{0-headway distributions in the generic case: (a) STS regime with $p_0=0.5$, $p=0.1$ and (b) FTS regime with $p_0=0.1$, $p=0.5$. iCOMF results for $P_{00}(0)$ ($P_{10}(0)$) is shown in blue (red) continuous lines. They are compared with computer simulations represented with markers.}
    \label{fig:0headway-generic}
\end{figure}

\begin{figure}
    \centering
    \includegraphics[width=1\linewidth]{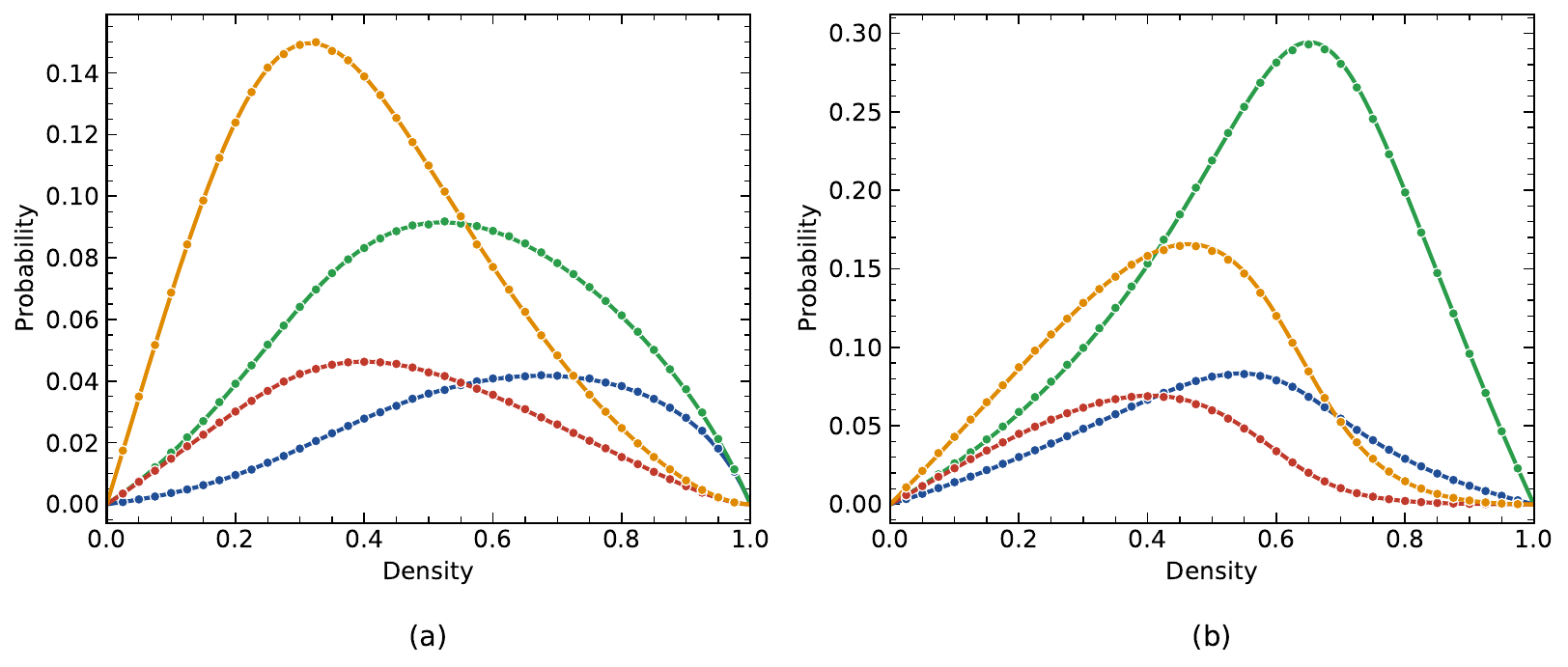}
    \caption{1-headway distributions in the generic case: (a) STS regime with $p_0=0.5$, $p=0.1$ and (b) FTS regime with $p_0=0.1$, $p=0.5$. iCOMF results for $P_{00}(1),P_{10}(1), P_{01}(1), P_{11}(1)$ are shown as continuous blue, red, green and orange, respectively, and are compared with computer simulations (markers).}
    \label{fig:Generic_headway1}
\end{figure}

Unlike the cruise-control limit, where COMF seems to produce the exact fundamental diagram despite failing to describe the headway distributions in the congested regime, for the case $p>0$ this "error cancellation" no longer holds and the fundamental diagram given by COMF deviates from the results of computer simulations (Fig.~\ref{FD- General Model}), especially around the maximal current density. In the STS regime, due to tendency towards phase separation, the microscopic configurations are inhomogeneous. For instance, at intermediate and high densities, a blocked stationary car is likely to have another such car in front. These correlations cannot be captured by COMF which underestimates zero-headway distributions, and consequently overestimates the flow. Taking into account the velocity of the car ahead, iCOMF accounts for the relevant correlations and shows excellent agreement with computer simulations  for the fundamental diagram (Fig.~\ref{FD- General Model}b) and also for the headway distributions as shown in Fig.~\ref{fig:0headway-generic}b and Fig.~\ref{fig:Generic_headway1}b. 
We note here that although it not a physically relevant regime, for the fast-to-start (FTS) regime $p_0 < p$, the headway distributions cannot be described by COMF either as the states remain inhomogeneous. For instance, computer simulations indicate that at high densities for the case $p_0=0$ and $p>0$, starting from a jammed configuration, the system evolves into a configuration consisting of a "mega jam" in the form "e10...0e"  and the rest of the lattice exhibits a repeating "e10e" pattern, i.e.\ a somewhat different phase separated state compared to the STS regime. In this regime, COMF underestimates the probability of a car joining a stopped vehicle ahead and underestimates the flux whereas iCOMF can adequately capture the correlations for intermediate and high densities as we observe excellent agreement between the iCOMF fundamental diagram and computer simulations (Fig.~\ref{FD- General Model}a). 

Finally, the headway distributions of the general model are also in excellent agreement with computer simulations as shown in Fig.~\ref{fig:0headway-generic} and Fig.~\ref{fig:Generic_headway1}. This suggests that iCOMF theory may even be an exact description of the VDR model with $v_{\text{max}}=1$.

\section{Summary and Discussion}

We have introduced an analytical method to describe the stationary states of single-lane stochastic cellular automaton models of traffic flow. 
The original COMF theory is a true mean-field theory for the headway $n$ of car with velocity $v$. It has previously been applied successfully to the NaSch model where it becomes exact for the case $v_\text{max}=1$. For $v_\text{max}>1$ or for its simple generalizations, such as the VDR model, even for $v_\text{max}=1$, the particle-hole symmetry is broken and COMF only yields approximate results. Here we have extended the approach by taking into account the velocity of the car immediately in front to capture additional correlations, and have exemplified this extension for the VDR model with $v_{\text{max}}=1$. This is done so to especially capture the correlations that are relevant to inhomogeneous steady states, such as the phase-separated configurations of the STS regime, which are relevant for traffic applications and are not well described by the original COMF.
    
The corresponding master equations of the iCOMF theory, are solved analytically using a generating function approach. The resulting fundamental diagram and headway distributions are in excellent agreement with computer simulations for all values of the model parameters $p_0$ and $p$ and across all density range, including intermediate and high densities dominated by inhomogeneous configurations. Therefore, unlike for the NaSch model with $v_{max}=1$, where the steady state factorizes and consecutive cars are uncorrelated, for the VDR model with $v_{max}=1$ the correlations between consecutive cars are relevant, especially at intermediate and high densities. The inhomogeneous states seem to appear due to correlation between a car's local headway and velocity configuration and the velocity of the car immediately ahead.

In the cruise-control limit of the VDR model with $v_{\text{max}}=1$ ($p=0$), where a moving car can only slow down due to insufficient number of empty sites ahead, we obtained explicit solutions of both the COMF and iCOMF master equations. In this limit, phase separation appears in its strongest form: a mega-jam coexisting with a free-flow region. Although both approaches reproduce the stable and metastable branches of the fundamental diagram, we have shown that the COMF results are only exact below the density $\frac{1-p_0}{2-p_0}$.  Above this density, the excellent agreement between the fundamental diagram obtained with the COMF and computer simulations is largely coincidental: the COMF headway distributions fail to capture the underlying phase-separated structure. By contrast, we have shown that iCOMF reproduces these inhomogeneous steady states.

It would be desirable to clarify, using iCOMF theory, whether metastable states exist in the general slow-to-start case where $0<p<p_0\leq 1$. In this regime, computer simulations do not indicate the presence of metastable states, but we believe this may be attributed to their very short lifetime due to fluctuations.
On the other hand, for $v_{\text{max}}>1$, computer simulations are able to capture metastable states and it is therefore also of considerable interest to apply the iCOMF theory for $v_{\text{max}}>1$ and examine whether metastable states of the VDR model with a longer interaction range can be described analytically.

In light of these arguments, the results of the iCOMF theory for the VDR model with $v_{\text{max}}=1$ may, in fact, be exact. We also note, the iCOMF theory yields excellent agreement for other models in the slow-to-start cellular automaton class (Takayasu–Takayasu \cite{TakayasuT93a} and Benjamin–Johnson–Hui \cite{BenjaminJH96} model), and even for the original NaSch models with longer interaction range $v_{\text{max}}>1$.  These findings will be presented in forthcoming publications \cite{iCOMF}.


\appendix

\section{Solution of iCOMF master equations}
\label{Appendix-iCOMFdetails}

We define $P_{uv}(n,t)$ the probability of finding exactly $n$ empty sites between a car with velocity $u$ and the car immediately ahead with velocity $v$. In the thermodynamic limit, the iCOMF master equation for the steady state are given by
\begin{align}
\label{iCOMF VDR}
P_{00}(0) &= \bar{g}_0 \left[ P_{00}(0) + P_{10}(0) \right], &&   \\
P_{00}(n) &= \bar{g}_0 \left[ p_0 P_{00}(n) + p P_{10}(n) \right] + \bar{g}_1 \left[ p_0P_{01}(n) + pP_{11}(n) \right],  \qquad \forall n \geq1,    \\
P_{01}(0)&=0, &&   \\
P_{01}(1) &= g_0(P_{00}(0) + P_{10}(0)), &&   \\
P_{01}(n) &= g_0 \left[ p_0P_{00}(n-1)+pP_{10}(n-1)\right] \nonumber \\&\hspace{3cm} + g_1\left[p_0P_{01}(n-1)+pP_{11}(n-1)\right], \qquad \forall n \geq 2, \\
P_{10}(n) &= \bar{g}_0\left[(1-p_0)P_{00}(n+1)+(1-p)P_{10}(n+1)\right] \nonumber\\ 
          &\hspace{3cm} +\bar{g}_1\left[(1-p_0)P_{01}(n+1)+(1-p)P_{11}(n+1)\right], \quad \forall  n \geq 0,  \\
P_{11}(0) &= 0, &&  \\
\label{iCOMF_VDR_P11(n)}
P_{11}(n) &= g_0\left[(1-p_0)P_{00}(n)+(1-p)P_{10}(n)\right] \nonumber \\&\hspace{3cm} +g_1\left[(1-p_0)P_{01}(n)+(1-p)P_{11}(n)\right], \quad \forall n \geq 1,
\end{align} 
where the conditional probabilities $g_v$ are given by \eqref{g_v}. 
We note that $P_{01}(0)$ and $P_{11}(0)$ both vanish due to representing Garden-of-Eden states  \cite{SchadschneiderS98}.

The system of equations \eqref{iCOMF VDR}--\eqref{iCOMF_VDR_P11(n)} can be solved by generating functions \cite{SchadschneiderS97}, we introduce
\begin{align}
\label{Def of F}
     F_{uv}(z)&=\sum_{n=0}^{\infty}P_{uv}(n)z^{n+1}, \qquad u,v \in \{0,1\}, \\
     \label{Total F}
     F(z) &= \sum_{u=0,1}\sum_{v=0,1} F_{uv}(z).
\end{align}
 For a fixed pair $u,v$ multiplying each equation in \eqref{iCOMF VDR}--\eqref{iCOMF_VDR_P11(n)} by $z^{n+1}$ and summing over all $n\geq0$ leads to a linear system of equations in $F_{uv}$ and we obtain the expressions:
\begin{align}
\label{eq:Solution_Gen_functions_VDR}
    F_{11}(z) &= \frac{g_0\big[g_1-(g_1-g_0)p_0\big]\big[(1-p_0)P_{00}(0)+(1-p)P_{10}(0)\big]z^2}{[\bar{g}_1+(g_1-g_0)p][1-\bar{g}_0 p_0-g_0 p]-z[g_1-(g_1-g_0)p_0][\bar{g}_1 p_0+g_1 p]} ,\\
    F_{10}(z) &= z^{-1}\frac{\bar{g}_1 +p(g_1-g_0)}{g_1-p_0 (g_1-g_0)} F_{11}(z), \\
    F_{01}(z) &= \frac{\bar{g}_1 +p(g_1-g_0)}{g_1-p_0 (g_1-g_0)} F_{11}(z) ,\\ \label{eq:F11}
    F_{00}(z) &= \frac{\bar{g}_0(\bar{g}_1 +p(g_1-g_0))-z(g_1-g_0)(p_0\bar{g}_1+p g_1)}{z g_0(g_1-p_0 (g_1-g_0))} F_{11}(z).
\end{align}

We define the shorthand notations $a=P_{00}(0)$ and $b=P_{10}(0)$. Using \eqref{g_v} and the fact that the conditions
\begin{align}
    F(1) &= 1, & P_0 &= F_{00}(1) + F_{01}(1),
\end{align}
must be satisfied, we find
\begin{align}
    P_0 &= \frac{p+(1-p_0)a + (1-p)b}{1 + p - p_0},\\
    \label{eq:g0_intermsof_ab}
    g_0 &= (1-p_0) \frac{p(1-a)+(1-p)b}{p+(1-p_0)a+(1-p)b},\\ 
    g_1 &= (1-p) \frac{(1-p_0)(a-1) + (2-p_0)b}{(1-p_0)(a-1)+(1-p) b}.
\end{align}
The generating functions \eqref{eq:Solution_Gen_functions_VDR}--\eqref{eq:F11} can now be expressed solely in terms of the two free variables $a$ and $b$. 

Fixing $a$ and $b$ requires two independent relations. Let $s=a+b$ be the total zero-headway probability. Using \eqref{iCOMF VDR} we have
\begin{align}
\label{eq:a_b_transformation}
    a &= s(1-g_0), & b & =s g_0,
\end{align}
and the equation \eqref{eq:g0_intermsof_ab} becomes
\begin{equation}
\label{eq:total_zero_headway}
    s = \frac{p(\bar{g}_0-p_0)}{p(1-p_0)+g_0^2(p_0-p)}.
\end{equation}
The last relation is given by the density condition $F'(1)=\frac{1}{\rho}$ and leads to 

\begin{equation}
\label{eq:cubic}
\begin{aligned}
0={}&(2\rho-1)(p_0-p)g_0^3
\\
&+\left[
(1-p_0)(p_0-p)
+\rho\left(p_0^2-2p_0p-2p_0+3p\right)
\right]g_0^2
\\
&-(1-p_0)pg_0
+(1-\rho)p(1-p_0)^2 .
\end{aligned}
\end{equation}
Therefore, the problem reduces to finding the roots $g_0$ of this cubic, from which the zero-headway probabilities follow via \eqref{eq:a_b_transformation}--\eqref{eq:total_zero_headway}, along with all remaining quantities.

The headway distributions $P_{11}(n)$, for all $n\geq 1$ are given by
\begin{equation}
    P_{11}(n) = \frac{g_0\bigl[a(1-p_0) + b(1-p)\bigr]}{\bar{g}_1 p_0 + g_1 p}\left[\frac{(g_1 - (g_1-g_0)p_0 )(\bar{g}_1 p_0 + g_1 p)}{(\bar{g}_1 + (g_1-g_0) p)(1 - \bar{g}_0 p_0 - g_0 p)}\right]^{n}.
\end{equation}
All the remaining ones for $n\geq0$, follow as
\begin{align}
    P_{01}(n) &= \frac{\bar{g}_1 + p(g_1-g_0)}{g_1-p_0(g_1-g_0)}P_{11}(n) ,  \\
    P_{10}(n) &= \frac{\bar{g}_1 + p(g_1-g_0)}{g_1-p_0(g_1-g_0)}P_{11}(n+1) , \\
     P_{00}(n) &= \frac{\bar{g}_0\big(\bar{g}_1 + p(g_1-g_0)\big) P_{11}(n+1)-(g_1-g_0)(p_0\bar{g}_1+pg_1) P_{11}(n)}{g_0(g_1-p_0(g_1-g_0))}. 
\end{align}
Finally, the fundamental diagram is given by
\begin{equation}
\label{eq:iCOMF_Current}
    J(c) = \rho\sum_{v=0,1} g_v P_v = \rho\Big(1-p+ (p-p_0)P_0 -(1-p_0)a - (1-p)b \Big).
\end{equation}


\section{Solution of COMF master equations }\label{sec: Original COMF}

We define $P_v(n,t)$ the probability of finding exactly $n$ empty sites in front of a car with velocity $v$ at time $t$. In the thermodynamic limit, the COMF master equation for the steady state are given by \cite{Barlovic-Diploma}:
\begin{align}
\label{COMFVDR}
P_0(0) &= \bar{g} \left[ P_0(0) + P_1(0) \right], \\
P_0(1) &= \bar{g} \left[ p_0 P_0(1) + p P_1(1) \right] + g \left[ P_0(0) + P_1(0) \right],  \\
P_0(n) &= \bar{g} \left[ p_0 P_0(n) + p P_1(n) \right] + g \left[ p_0 P_0(n-1) + p P_1(n-1) \right], \qquad\quad \forall n \geq 2, \\
P_1(0) &= \bar{g} \left[ (1-p_0) P_0(1) + (1-p) P_1(1) \right],\\
\label{COMFVDR_P1(n)}
P_1(n) &= \bar{g} \left[ (1-p_0) P_0(n+1) + (1-p) 
P_1(n+1) \right] \nonumber \\ 
&\hspace{4.1cm}  + g \left[ (1-p_0) P_0(n) + (1-p) P_1(n) \right], \qquad \forall  n \geq 1,
\end{align}
where 
\begin{equation}
    g=\sum_{v=0,1}(1-p(v))\sum_{n=1}^{\infty}P_v(n)
\end{equation} is the probability for a car to move one site. The system of equations \eqref{COMFVDR}--\eqref{COMFVDR_P1(n)} can be solved similarly to \ref{Appendix-iCOMFdetails}. For $v=0,1$ we define
\begin{align}
\label{eq: Generating Function COMF - Generic}
    F_v(z) &= \sum_{n=0}^\infty z^{n+1}P_v(n), \\
    F(z) &=F_1(z) + F_0(z)=\sum_{n=0}^\infty z^{n+1}P(n),
\end{align}
and obtain
\begin{align}
\label{eq:COMF Generating}
F_0(z) &= \frac{\bar{g}z(\bar{g}+gz)\Big((1-p_0)P_0(0) + (1-p)P_1(0)\Big)}{(\bar{g}+gz)(1-p_0\bar{g}-pg) - gz}, \\
F_1(z) &= \frac{g}{\bar{g}}F_0(z),
\end{align}
with $\bar{g}=1-g$. The conservation of probability i.e. $F(z=1)=1$ leads to
\begin{equation}
\label{eq:g explicit for COMF}
    g = \frac{(1-p_0)\big(1-P_0(0)\big) -(1-p) P_1(0)}{1+p-p_0}.
 \end{equation}

The problem now reduces to determining $P_0(0)$ and $P_1(0)$ which requires two relations between these quantities. The first of these is the global density condition $F'(z=1)=\frac{1}{\rho}$ and reads:
\begin{equation}\label{eq:COMF_density}
\frac{[P_0(0)(1-p_0)+P_1(0)(1-p)][\bar{g}(1+g-p_0)-g p]}{[\bar{g}(1-p_0)-g p]^2}=\frac{1}{\rho}.
\end{equation}
The second is simply the equation \eqref{COMFVDR}. Using \eqref{eq:g explicit for COMF} these two equations become quadratic in $P_0(0), P_1(0)$ and have to be solved simultaneously. Using $P(0)=P_0(0) + P_1(0)$, the latter can be expressed as
\begin{align}
\label{eq: P_0(0)}
P_0(0) &=
\frac{
P(0)\bigl(p+P(0)(1-p)\bigr)
}{
1+p-p_0+P(0)(p_0-p)
},
\\[6pt]
P_1(0) &= P(0)-P_0(0)=
\frac{
P(0)(1-p_0)(1-P(0))
}{
1+p-p_0+P(0)(p_0-p)
},
\end{align}
substituting in \eqref{eq:COMF_density} leads to the quadratic relation for $P(0)$
\begin{equation}
\label{eq:quadratic P(0)}
    \big[-\rho+p_0(\rho-1)+p\big]P(0)^2 +\big[(2-p-p_0)c-1+p_0-p\big]P(0) + p\rho = 0 .
\end{equation}

Solving \eqref{eq:quadratic P(0)} and, using \eqref{eq: P_0(0)}, the headway distributions for $n\geq1$ is obtained through
\begin{align}
    P_0(n) &= \frac{g[(1-p_0) P_0(0) + (1-p) P_1(0)]}{[(1-p_0)\bar{g} + (1-p) g]^2}\left[\frac{g}{\bar{g}}\,\frac{p_0\bar{g} + p g}{(1-p_0)\bar{g} + (1-p) g}\right]^{n-1}, \\
    P_1(n) &= \frac{g}{\bar{g}}P_0(n).
\end{align}

The fundamental diagram can be explicitly calculated using $g = \frac{P_1(0)}{P(0)}$ and re-writing  \eqref{eq:quadratic P(0)} as
\begin{equation}
\label{equation g}
    \rho g^2-\big(1+(p_0-p)(\rho-1)\big)g-(1-p_0)(\rho-1)=0,
\end{equation}
from which we obtain
\begin{align}
\label{comf current}
    J_{\pm}(\rho)= \rho g_{\pm} =\frac{1}{2}\bigg(1+(p_0-p)(\rho-1) \pm \sqrt{(1+(p_0-p)(\rho-1))^2+4\rho(1-p_0)(\rho-1)}\bigg). 
\end{align}



\section*{Artificial Intelligence Disclosure Statement}

During the preparation of this manuscript, we acknowledge the use of
Large Language Models (LLMs), namely Claude (models: Opus 4 and 5, Anthropic) and ChatGPT (models: GPT-4 and GPT-5, OpenAI).
These assisted with improving clarity of some parts of the text through grammar and spelling corrections, with developing and debugging codes for computer simulations, and with checking parts of the analytical calculations
already carried out by the authors. The authors take full responsibility and ownership of the work presented in this paper.
\section*{References}
\bibliographystyle{unsrt}
\bibliography{references}

\end{document}